\documentclass[a4paper,11pt]{article}
\usepackage{jinstpub} 
\usepackage{color,soul}

\proceeding{TWEPP 2024 Topical Workshop on Electronics for Particle Physics\\
  Sep 30 – Oct 4, 2024\\
  Glasgow, Scotland, UK}

\title{\boldmath An Integer-$N$ Frequency Synthesizer for Flexible On-Chip Clock Generation}

\author[1]{S. Mandal,\note{Corresponding author.}}
\author{P. Maj,}
\author{G. W. Deptuch}

\affiliation{Instrumentation Department, Brookhaven National Laboratory,\\
Upton, NY 11973-5000, USA}

\emailAdd{smandal@bnl.gov}

\abstract{A low-power integer-$N$ frequency synthesizer for flexible on-chip clock generation has been designed in a 65\,nm CMOS process. The circuit can be programmed to generate two independent low-jitter clocks between 30\,MHz and 3\,GHz that are locked to a 10-50\,MHz reference input. The design uses a phase-locked loop (PLL) with a dual-tuned $LC$ voltage-controlled oscillator (VCO), programmable feedback divider, and dual output dividers. The total power consumption from 1.2\,V and 0.8\,V supplies is 4.0\,mW. Experimental results confirm the functionality of the proposed synthesizer over a wide range of output frequencies.}


\arxivnumber{2411.01552} 

\begin{document}
\maketitle
\flushbottom

\section{Introduction}
\label{sec:intro}

High-energy physics experiments aiming  at detection of rare events often require high-data-rate links between readout ASICs and back-end processors over lossy channels, such as radio-pure cables~\cite{capozzi2019dune,adhikari2021nexo,albert2018sensitivity}. For example, the nEXO experiment~\cite{adhikari2021nexo} will require $\sim$400 radio-pure data links, each operating at speeds of about 500\,Mb/s in a liquid xenon cryostat (165\,K). The cryogenic environment also imposes a strict power constraint of $\sim$15\,mW per link~\cite{john2024testing}. Proposed link designs for such challenging scenarios use the \emph{forwarded clock architecture} shown in Fig.~\ref{fig:PLL_top}(b). In this approach, the data acquisition (DAQ) system transfers a low-frequency reference clock (e.g., from a crystal oscillator) to the analog front-end (AFE) and readout system over a low-speed cable, which is then used by an on-chip frequency synthesizer~\cite{biereigel2020sissa} to generate the high-frequency clock required for data serialization. Here we describe a low-power integer-$N$ frequency synthesizer with a wide output frequency range suitable for such clock generation applications.


\section{Circuit Design}
\label{sec:circuit_design}

\begin{figure}
    \centering
    \includegraphics[width=0.36\textwidth]{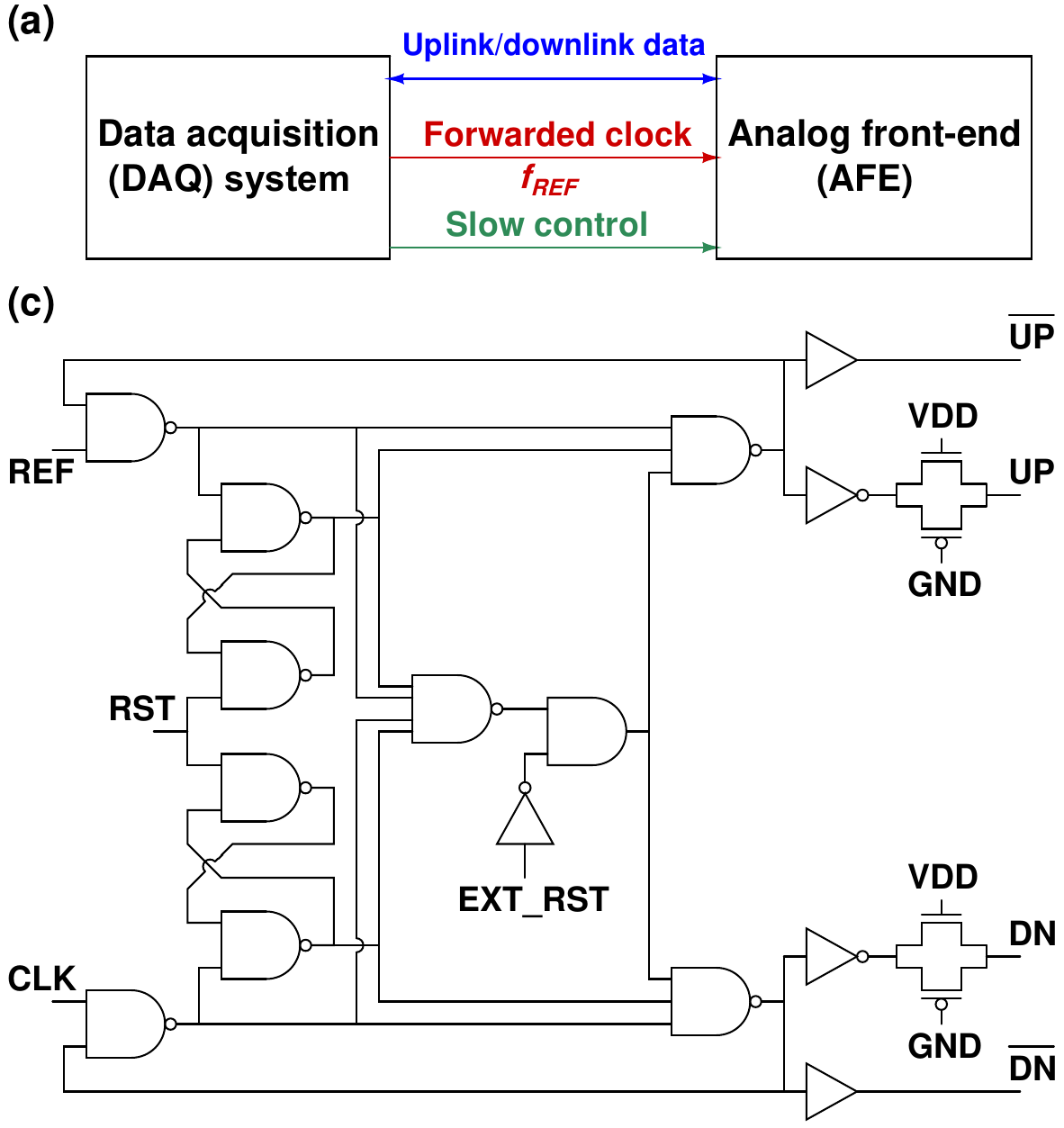}~
    \includegraphics[width=0.62\linewidth]{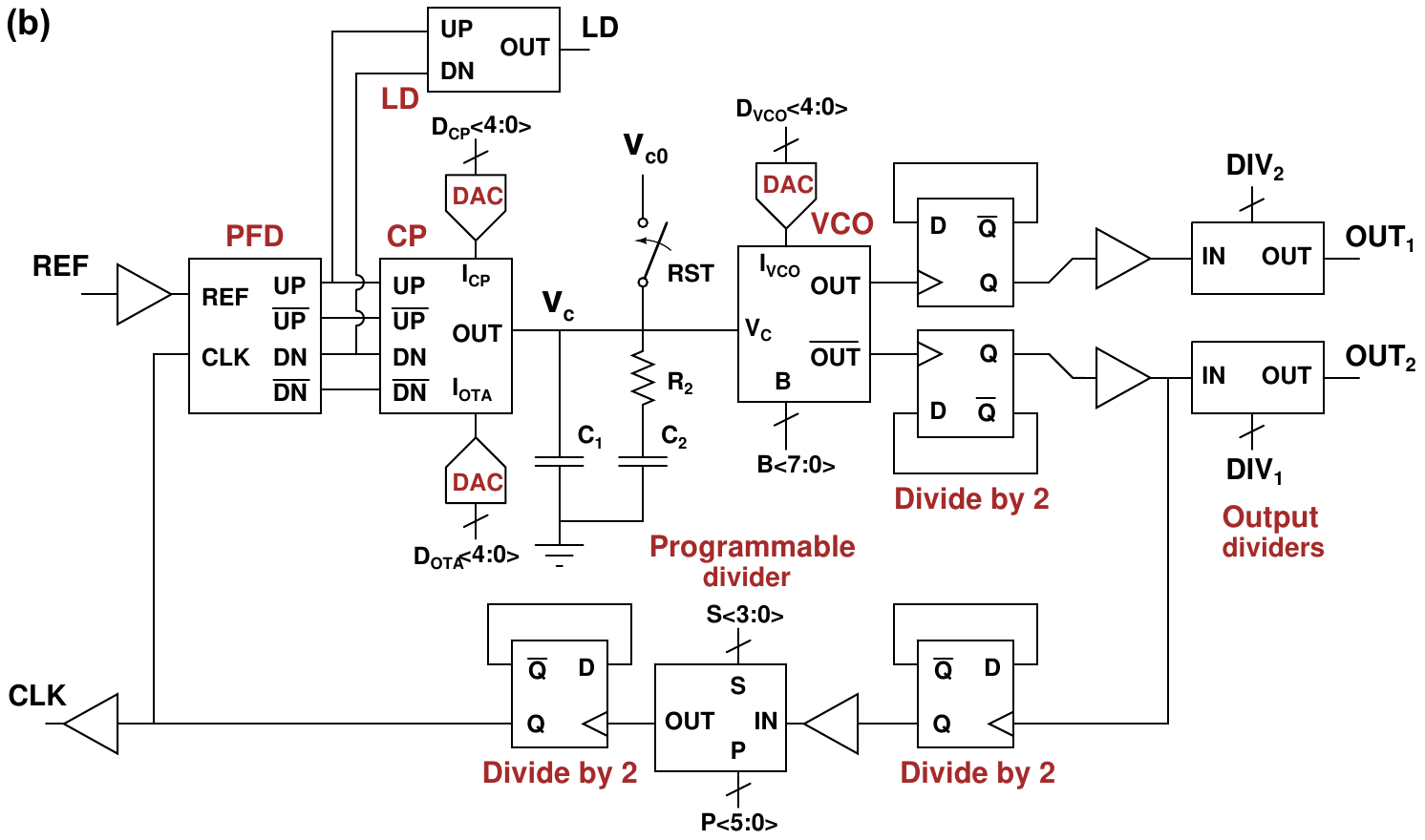}
    \caption{(a) Overview of the forwarded clock architecture. (b) Top-level block diagram of the proposed frequency synthesizer. (c) Schematic of the PFD including reset signals (RST and EXT\_RST).}
    \label{fig:PLL_top}
\end{figure}

Fig.~\ref{fig:PLL_top}(b) shows a top-level block diagram of the proposed synthesizer. The design uses a phase-locked loop (PLL) with programmable feedback divider and dual output dividers to allow generation of two independent output frequencies. The PLL has a classical structure with an edge-triggered phase-frequency detector (PFD), differential charge pump (CP), passive $RC$ loop filter, voltage-controlled oscillator (VCO), frequency divider, and lock detector (LD). All required biases  are generated on-chip via 5-bit current DACs, resulting in a completely self-contained design.

The PFD, which is shown in Fig.~\ref{fig:PLL_top}(c), is based on the design in~\cite{young1992pll}. It is an efficient implementation of the commonly-used sequential PFD circuit using only NAND gates and inverters. The design is 1) insensitive to input duty cycle; 2) guarantees a minimum up/down (UP/DN) output pulse width in the locked state, thus eliminating any dead zone around the origin of the PFD transfer curve; and 3) has a relatively small delay (three NAND gates) in the critical path.

\begin{figure}
    \centering
    \includegraphics[width=1\linewidth]{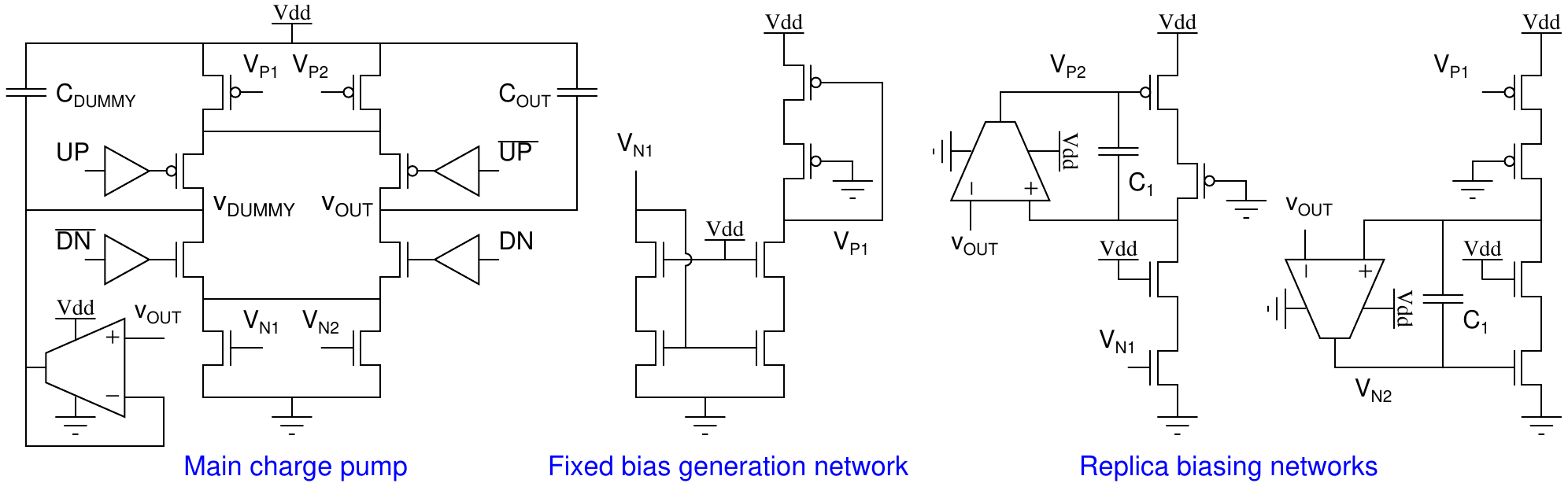}
    \caption{Schematic of the CP, including both the fixed and adaptive biasing networks.}
    \label{fig:charge_pump_3_rev}
\end{figure}

The CP is shown in Fig.~\ref{fig:charge_pump_3_rev}. The design uses differential switching to minimize charge injection. An OTA maintains the dummy output voltage ($v_{DUMMY}$) near the actual output voltage ($v_{OUT}$) to minimize switching transients. Additionally, a replica-bias loop is used to improve matching between the P- and N-sides, thus reducing phase offset in the locked state. For this purpose, both the N- and P-side bias transistors are split into two equal parts. The gate voltages for one part ($V_{N1}$ and $V_{P1}$) are set by a fixed input current, while those for the other part  ($V_{N2}$ and $V_{P2}$) are adaptively set by two replica biasing networks as shown in the Fig.~\ref{fig:charge_pump_3_rev}.

\begin{figure}
    \centering
    \includegraphics[width=0.58\linewidth]{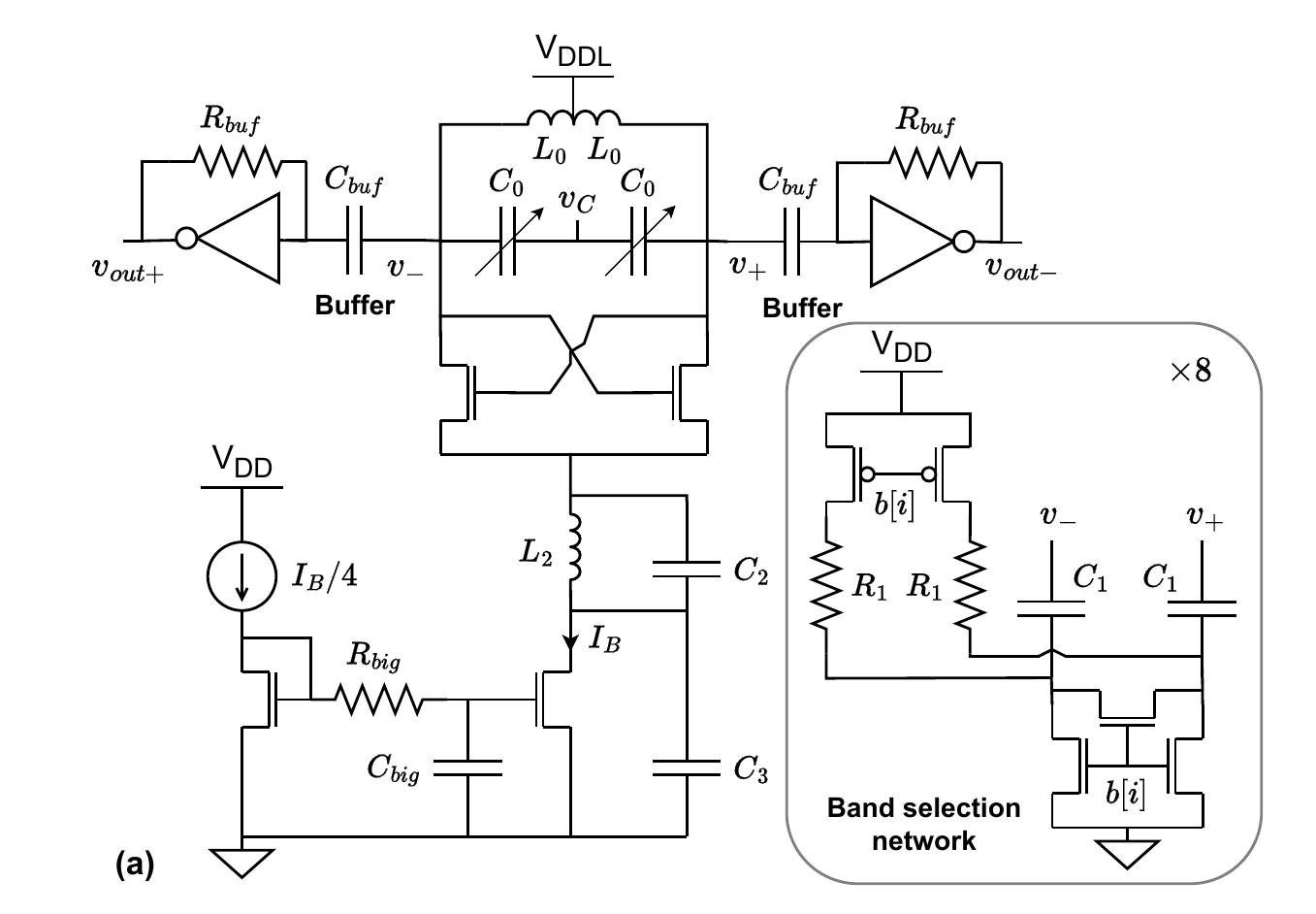}
    \includegraphics[width=0.30\linewidth]{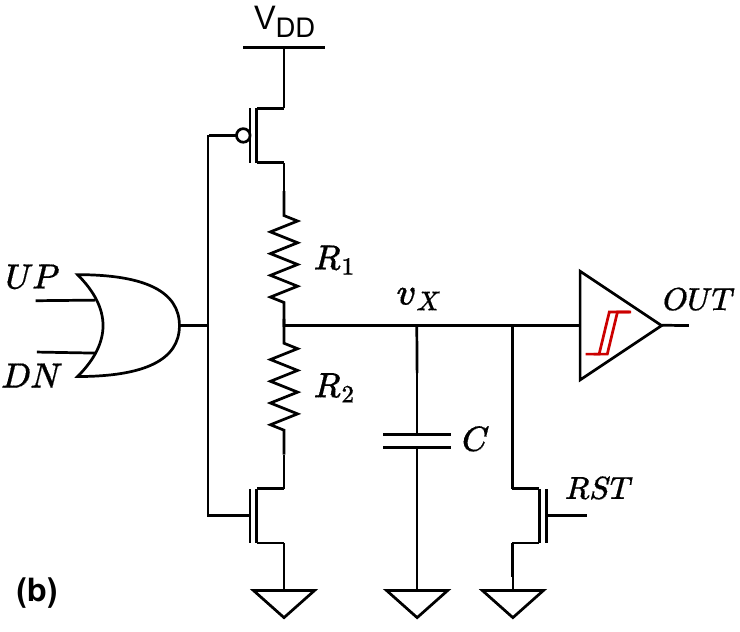}
    \includegraphics[width=0.08\linewidth]{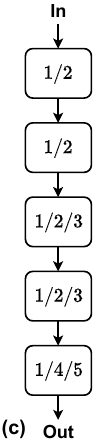}
    \caption{(a) Schematic of the VCO, including both coarse- and fine-tuning networks. (b) Schematic of the lock detector (LD). (c) Block diagram of the output divider.}
    \label{fig:VCO_schematic_rev}
\end{figure}

The VCO, which is shown in Fig.~\ref{fig:VCO_schematic_rev}(a), uses an on-chip $LC$ resonator to minimize phase noise and power supply sensitivity. The inductor is a center-tapped circular spiral structure, while positive feedback is provided by a cross-coupled NMOS pair. Two levels of frequency tuning are available: coarse tuning (band switching) using 8 equal-valued switched capacitors, and fine tuning using accumulation-mode MOS varactors. The coarse tuning network uses differential switching, as shown in the figure, to minimize the impact of switch on-resistance on the tank's quality factor, $Q$. The switching network includes 1) small NMOS switches to ground that define the $V_{DS}$ of the main switch when it is ``on'', and 2) small PMOS switches and series resistors to $V_{DD}$ that prevent $V_{DS}$ from exceeding $V_{DD}$ when it is ``off''. Phase noise is further minimized by using 1) an $RC$ low-pass filter to remove bias current noise, and 2) an $LC$ tail current filter (tuned to the second harmonic) to reduce $1/f$ noise up-conversion~\cite{hegazi2001filtering}. Power consumption is minimized by operating the VCO core at a reduced supply voltage, $V_{DDL}$. The simulated tuning range for $V_{DDL}=0.8$\,V is 5.6-8.6 GHz with a control gain of $\sim$0.6\,GHz/V and a typical phase noise of $-110$\,dBc/Hz at 1\,MHz offset. 

The VCO outputs are buffered to logic-level signals by AC-coupled CMOS inverters with resistive feedback, as shown in Fig.~\ref{fig:VCO_schematic_rev}(a). While this design ensures high gain and insensitivity to the VCO's output common-mode voltage, it also degrades supply sensitivity since the supply-dependent input capacitance of the inverters contributes to the total capacitance of the $LC$ resonator. This effect can be minimized by running the buffers off a separate low-noise power supply.

The feedback divider uses a pulse-swallow counter with a total division ratio of $(NP+S)$, where $N=2$ is fixed, while $P$ and $S$ are adjustable over the ranges 1-63 and 0-15, respectively. Propagation delays are minimized by using dynamic true single-phase clocking (TSPC) flip-flops within the $N/(N + 1)$ dual-modulus prescaler and a one-level carry select adder (CSA) within the programmable counter. The divider output is further divided by 2 to ensure that the recovered clock (CLK) has a 50\% duty cycle. The second-order passive $RC$ loop filter was optimized for a reference frequency of $f_{REF}=50$\,MHz, a charge pump current of 10\,$\mu$A, and a loop bandwidth of 230\,kHz.

Fig.~\ref{fig:VCO_schematic_rev}(b) shows a schematic of the lock detector (LD) circuit, which detects when total time-averaged duty cycle of the PFD's UP and DN outputs, $D$, is less than a threshold value, $D_{min}$. In steady-state, charge balance at the node $v_{x}$ requires $v_{x} = V_{DD}\left(\frac{D}{1-D}\frac{R_1}{R_2}+1\right)^{-1}$. The output of the Schmitt trigger goes high when $v_{x}\geq \alpha V_{DD}$, its positive-going threshold voltage. Thus, the circuit indicates ``lock achieved'' when $D\leq D_{min}$, where $D_{min}=\left(\frac{\alpha}{1-\alpha}\frac{R_1}{R_2}+1\right)^{-1}$. Note that $D_{min}$ is set by two dimensionless quantities, namely $\alpha$ and $R_1/R_2$, and is thus robust to changes in process, voltage, and temperature (PVT). For this design, $\alpha\approx 2/3$ and $R_1/R_2 = 8$, resulting in $D_{min}\approx 0.059$.

Fig.~\ref{fig:VCO_schematic_rev}(c) shows a block diagram of the programmable output divider. It uses a cascade of selectable divider stages with decreasing division ratios, as shown. Each stage also includes a ``divide by 1'' mode, which is implemented by using a transmission gate multiplexer to bypass the divider circuit. Overall, the design allows a wide range of divide values from 1-160 (including most composite numbers $<32$) to be generated while maintaining an output duty cycle close to 50\%. 




The complete synthesizer uses dual output dividers to allow generation of two programmable output frequencies from the quadrature outputs of the PLL and a standard two-wire I$^2$C serial interface for programming. The overall die area is 450\,$\mu$m\,$\times$\,500\,$\mu$m, of which approximately 50\% and 25\% are occupied by the VCO and main loop filter capacitor, respectively. The nominal power consumption (for $V_{DD}=1.2$\,V and $V_{DDL}=0.8$\,V) is 4.0\,mW. As an example, transient noise simulations for $f_{REF}=40$\,MHz and $f_{OUT}=1.92$\,GHz show a bimodal time interval error (TIE) distribution with deterministic and random jitter components~\cite{galton2018understanding} of 3.13\,ps and 560\,fs$_{rms}$, respectively; the former arises from mismatched rise and fall times of the CMOS output buffers.

\section{Experimental Results}
\begin{figure}[htbp]
\centering
\includegraphics[width=0.29\textwidth]{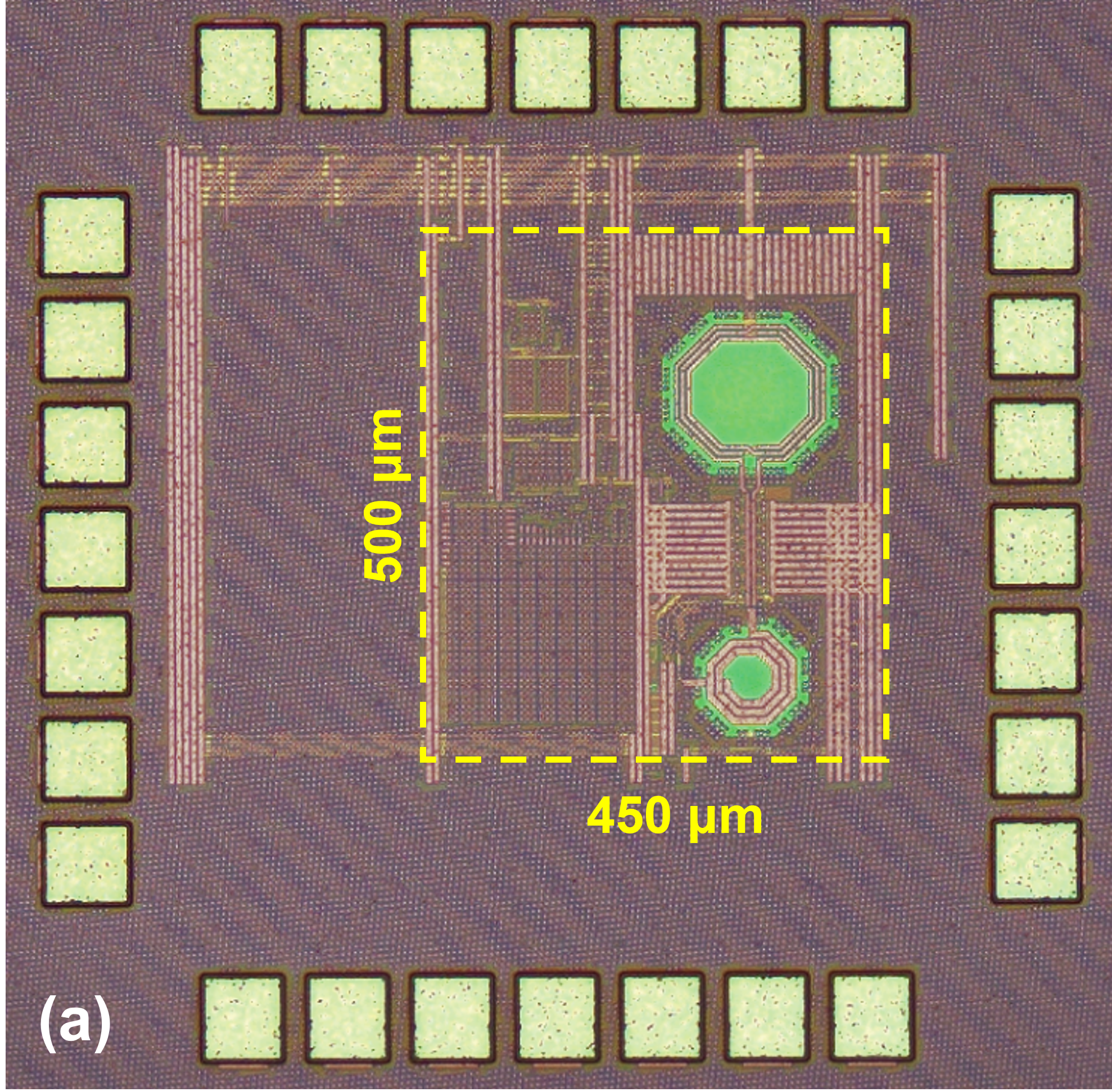}
\includegraphics[width=0.28\textwidth]{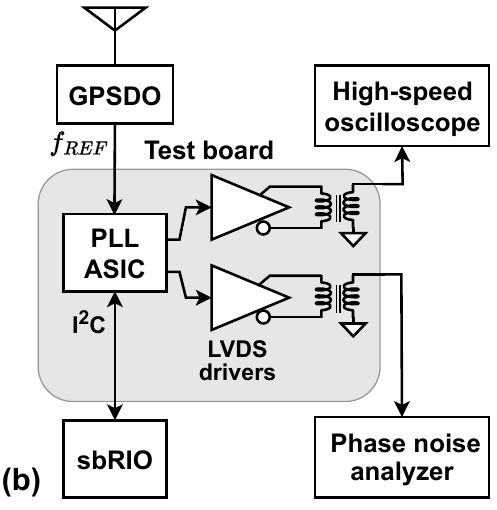}
\includegraphics[width=0.36\textwidth]{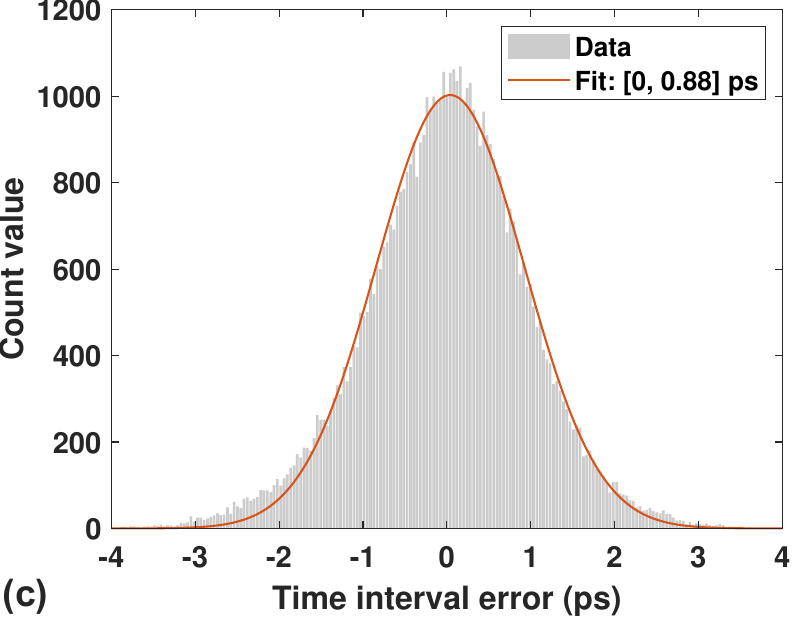}
\caption{(a) Die photograph of the fabricated test chip. (b) Block diagram of the experimental setup. (c) Measured time interval error (TIE) distribution at an output frequency of 1.92\,GHz for $f_{REF}=40$\,MHz.}
\label{fig:die_photo}
\end{figure}

Fig.~\ref{fig:die_photo}(a) shows a die photograph of the fabricated synthesizer ASIC. Due to a shortage of bond pads, all inputs and outputs were encoded as single-ended CMOS signals. Fig.~\ref{fig:die_photo}(b) summarizes the experimental setup used for room-temperature testing. The test board integrates the ASIC with 1) single-ended to LVDS drivers and baluns to generate impedance-matched outputs; 2) level translators for the I$^2$C interface; and 3) low-dropout linear voltage regulators (LDOs) for the main and VCO power supplies. The reference input is generated by a GPS-disciplined oscillator (GPSDO). The chosen device (LBE-1420, Leo Bodnar Electronics) has an internal PLL, thus allowing $f_{REF}$ to be easily varied during testing. The I$^2$C parameters are set by a graphical user interface (GUI) implemented on a single-board controller (sbRIO-9629 from National Instruments).

The effects of power supply noise and ripple on VCO stability can be analyzed by treating the power supply nodes ($V_{DD}$ and $V_{DDL}$) as sources of frequency pulling and estimating the resulting control gains, $K_{VDD}$ and $K_{VDDL}$. For the proposed design, simulations show $K_{VDD}\approx 48$\,MHz/V and $K_{VDDL}\approx 380$\,MHz/V, respectively, so we focus on the effects of $V_{DDL}$. The amplitude of the spur generated by a sinusoidal ripple, $v(t)=V_{m}\sin(2\pi f_m t)$, within $V_{VDDL}$ can be shown to be ${S_{spur}} = 10{\log _{10}}{\left( {\frac{{{K_{VDDL}}{V_m}}}{{2{f_m}}}} \right)^2}$. This result allows us to estimate the power supply rejection ratio (PSRR) requirements for the LDO. For example, keeping $S_{spur}<-60$\,dBc for a 10\,mV input ripple at $f_{m}=10$\,MHz requires PSRR $>36$\,dB. Similarly, an analytical formula for the effects of supply noise on phase noise (PN) can be obtained by assuming that only white noise sources are significant~\cite{urso2020analysis}, which is valid in the $1/f^2$ region of the PN spectrum. In this region, the supply-induced PN component is given by 
${{\cal L}_{sup}}(\Delta f) = 10{\log _{10}}\left( {\overline {v_{n,sup}^2(\Delta f)} {{\left( {\frac{{{K_{VDDL}}}}{{\Delta f}}} \right)}^2}} \right)$,
where $\overline{v^{2}_{n,sup}}$ is the power spectral density (PSD) of supply noise and $\Delta f$ is the offset frequency. Given the VCO's typical phase noise of ${\cal{L}}(\Delta f)\approx -110$\,dBc/Hz at $\Delta f=1$\,MHz, we need ${v_{n,sup}}<7.0$\,nV/Hz$^{1/2}$ to keep ${{\cal L}_{sup}}(\Delta f) < {{\cal L}(\Delta f)} $, thus limiting degradation of PN to $<3$\,dB. Accordingly, the board uses one of the few commercial LDOs with low-enough output noise PSD to meet this requirement (Analog Devices LT3042, $v_{n}\approx 2.0$\,nV/Hz$^{1/2}$ above 200\,Hz).

During initial tests, we set $f_{REF}=40$\,MHz and output divider values of 2 and 8, resulting in output frequencies of 1.92\,GHz and 480\,MHz, respectively. Eye diagrams and TIE values were measured using the built-in functions provided by a high-speed oscilloscope (Tektronix MSO58, 2\,GHz bandwidth). Fig.~\ref{fig:die_photo}(c) shows the measured TIE of the 1.92\,GHz output for nominal bias settings. The distribution is well fit by a zero-mean Gaussian, which suggests that it is dominated by random jitter. However, its estimated standard deviation of $\sim$880\,fs is degraded by the jitter noise floor (JNF) of the oscilloscope. Since the two jitter sources are uncorrelated, removing the estimated JNF of 540\,fs$_{rms}$ from the measurement yields an estimated output jitter of $\sqrt{880^2-540^2}\approx 700$\,fs$_{rms}$, which is in good agreement with the simulation results in Section~\ref{sec:circuit_design}.

\begin{figure}
    \centering
    \includegraphics[width=0.40\linewidth]{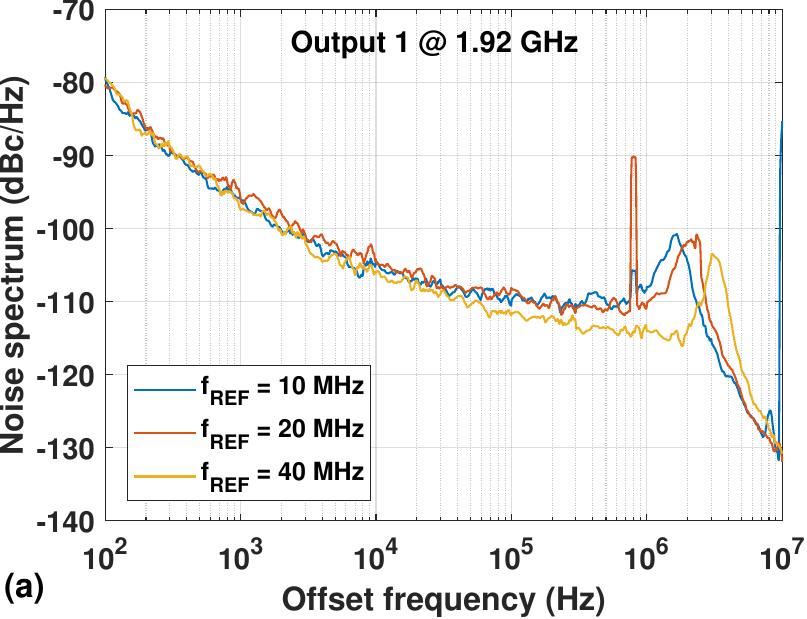}
    \includegraphics[width=0.40\linewidth]{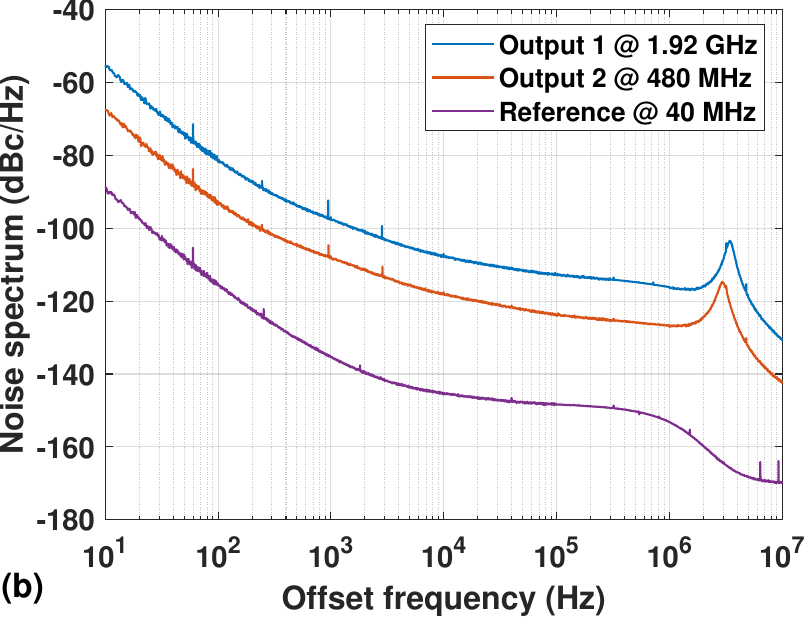}
    \caption{(a) Measured output phase noise (using the spectrum analyzer) at 1.92\,GHz for different values of $f_{REF}$. (b) Measured reference and output phase noise (using the phase noise analyzer) at $f_{REF}=40$\,MHz.}
    \label{fig:phase_noise}
\end{figure}

Phase noise measurements were performed using either 1) the phase noise mode of an RF spectrum analyzer (Rohde \& Schwarz FSW26), or 2) a dedicated phase noise analyzer (Rohde \& Schwarz FSWP50). The latter uses the cross-spectrum method to suppress internal noise sources, resulting in a significantly lower instrument noise floor for a given total measurement time~\cite{rubiola2010cross}. Fig.~\ref{fig:phase_noise}(a) shows the measured phase noise PSD at 1.92\,GHz for three different reference frequencies (10, 20, and 40\,MHz). As expected, increases in $f_{REF}$ result in 1) increased loop gain and closed-loop bandwidth; and 2) reduced multiplication of reference phase noise. Fig.~\ref{fig:phase_noise}(b) compares the measured PSDs of the reference and both outputs for a fixed $f_{REF}=40$\,MHz.

\section{Conclusion}
A low-power integer-$N$ synthesizer suitable for flexible on-chip clock generation has been designed in a 65\,nm CMOS process and tested at room temperature (300K). The measured jitter of 700\,fs$_{rms}$ enables downlink data transmission up to 3\,Gb/s. Future tests will focus on cryogenic operation. For this purpose, the test board will be modified to remove the LDOs, which are unlikely to function at cryogenic temperatures. Necessary voltage regulators will instead be placed outside the cryostat.




\bibliographystyle{JHEP}
\bibliography{biblio.bib}







\end{document}